# Efficient composite pulse sequences for arbitrarily accurate *z*-axis rotation gates

Li Zhang and Shihao Zhang

Beijing Key Laboratory of Nanophotonics and Ultrafine Optoelectronic Systems, School of Physics, Beijing Institute of Technology, 100081, Beijing, China

E-mail: 15201609482 @163.com

## Abstract

We propose various composite $\pi$-pulse sequences for implementing robust *z*-axis rotation gates widely used in quantum information processing (QIP) scenarios, and discuss their error tolerance of the pulse strength error (PSE) and off-resonance error (ORE), i.e., two typical systematic errors in NMR systems. Compared to existing composite gates, our designed composite pulses exhibit higher gate fidelity and shorter sequence time cost. Furthermore, such short pulse sequences are generalized by simple analytic formulas which can suppress these errors to any desired order simultaneously with linear length scaling, indicating an efficient construction of them with arbitrary accuracy.

(Some figures may appear in colour only in the online journal)

## 1. Introduction

In the past decades, the area of quantum computation (QC) and quantum information processing (QIP) has grown rapidly. With considerable progress made on technically improving the experimental conditions and operational levels for years [1-4], novel theoretical strategies are also desired for advancing practical QC and QIP architectures [5-8]. Concretely speaking, to perform accurate unitary transformations that are robust against realistic errors, a great deal of efforts from various aspects, including the fault-tolerant logic operations [9, 10], geometric [11, 12] or holonomic quantum computation [13-15], and topological error-cerrection [16, 17], have been devoted.

Among these developments in QIP, composite pulses have been devised to be an effective approach adapted from NMR scenarios to tackling particular systematic errors [18-25], especially the typical pulse strength errors (PSE) and off-resonance errors (ORE) denoted $\epsilon$ and *f*, respectively. For estimating the actually implemented composite pulses *V* with respect to the ideal unitary transformation *U*, the propagator fidelity $\mathcal{F} = |\mathrm{tr}(U^\dagger V)|/\mathrm{tr}(U^\dagger U)$ is usually adopted [26, 27]. In general, composite pulse sequences proposed with various number and types of constituent pulses would perform differently [24, 28, 29], and therefore how to implement desired robust unitary gates with high fidelity and shorter execution time becomes an inviting goal.

In this paper, we focus on the construction of arbitrary robust *z*-axis rotation gates denoted as

$$Z_\Phi = e^{-i\Phi/2}\begin{pmatrix} 1 & 0 \\ 0 & e^{i\Phi} \end{pmatrix}, \tag{1}$$

where $\Phi$ is the rotational angle about the *z*-axis of the Bloch sphere. Up to an unimportant global phase factor, the phase gate (denoted *S*) and $\pi/8$ gate (denoted *T*) are two specific instances of the $Z_\Phi$ gate class [30]

$$S = \begin{pmatrix} 1 & 0 \\ 0 & i \end{pmatrix} = Z_{\pi/2} \; ; \; T = \begin{pmatrix} 1 & 0 \\ 0 & e^{i\pi/4} \end{pmatrix} = Z_{\pi/4}. \tag{2}$$

These rotations together with Pauli-Z gate ($Z_\pi$) play an important role in QIP, such as for approximating quantum circuits to good accuracy [31, 32], quantum error-correction and fault-tolerant computation [33-35], and specific state distillation [36, 37]. Drawing ideas from several previous techniques, e.g., the geometric interpretation [27, 38], analytic or numerical approaches [20, 21, 39] and analysis of symmetric/antisymmetric properties [24, 25], we first investigate how to construct various composite pulses only consisting of $\pi$ rotations that can remove first-, second-, and third-order PSE and ORE in detail, and then compare these sequences with some existing schemes, which can reveal better performance of our design in some respects. Futhermore, we propose simple analytic formulas of such $\pi$-pulse sequences which are able to simultaneously suppress these two errors to any desired order $O(\epsilon^n)$ and $O(f^n)$ in the fidelities and their lengths scale linearly with $n$.

## 2. Robustness conditions

At first, the most basic condition that a $\pi$-pulse sequence must satisfy for the $Z_\Phi$ gate is the matrix product

$$\prod_{k=n}^{1}(\pi)_{\phi_k} = Z_\Phi, \quad \Phi = 2\sum_{j=1}^{n/2}(\phi_{2j} - \phi_{2j-1}), \tag{3}$$

where the index *k* descends from left to right and *n* is even. The notation $(\pi)_\phi$, or a general $(\theta)_\phi$ for a rotation by an angle $\theta$ around the $\phi$ axis in the *xy* plane is known as a propagator $(\theta)_\phi = \exp(-i\theta\sigma_\phi/2)$ with $\sigma_\phi = \cos(\phi)\sigma_x + \sin(\phi)\sigma_y$. The error PSE reflects the deviation of the strength of the driving field from its nominal value, which leads to some fraction $\epsilon$ in a single $\pi$ pulse as $(\pi)_\phi \to (\pi)_\phi(\epsilon\pi)_\phi$. Another error ORE occurs when driving field is not perfectly resonant with the qubit transtion and the off-resonance fraction *f* gives rise to $(\pi)_\phi \to \exp[-i\pi(\sigma_\phi + f\sigma_z)/2]$. Thus, the total propagator composed of *n* $\pi$ pulses is $V = \prod_{k=n}^{1}(\pi)_{\phi_k}(\epsilon\pi)_{\phi_k}$ or $V = \prod_{k=n}^{1}\exp[-i\pi(\sigma_{\phi_k} + f\sigma_z)/2]$ in the presence of PSE or ORE, respectively.

### 2.1. Baseline

We take $n=2$ in Eq. (3) to acquire a naive $Z_\Phi$ rotation as $Z_\Phi = (\pi)_0(\pi)_{-\Phi/2}$ without loss of generality, and such a rotation owns the baseline fidelity with respect to $\epsilon$ as

$$\mathcal{F}_\epsilon = 1 - \cos^2(\Phi/4)\pi^2\epsilon^2/2 + O(\epsilon^4) \tag{4}$$

and so has a second-order infidelity term.

Similarly, in the presence of ORE the baseline fidelity

$$\mathcal{F}_f = 1 - 2\sin^2(\Phi/4)f^2 + O(f^3) \tag{5}$$

is also only correct to second order. Notice that an infidelity of order $2n$ corresponds to an error term in the pulse of order $n$ as usual [24, 27]. The fidelities of the baseline solutions for $\Phi = \pi, \pi/2, \pi/4$ are presented in the first row of Fig. 1. In each plot, the ring-shaped region enclosed by the inmost contour line is quite narrow near the central point (especially along $\epsilon$) but can extend to certain large error values, e.g. $\mathcal{F}(\Phi = \pi/4, \epsilon = -1, f = -0.87) > 0.999$, leading to possible use in conventional NMR rather than in QC [24].

### 2.2. Composite pulse sequences

Suppressing higher-order error terms in the fidelity involves composite pulse sequences with even $n \geq 4$ in Eq. (3). To avoid repetition, here we directly refer to some well-established formulas and consequences from Ref. [27] with corresponding explanations.

In the presence of PSE, the $n$ $\pi$-pulse sequence can be arranged as $V = \prod_{k=n}^{1}(\pi)_{\phi_k}\prod_{k=n}^{1}(\epsilon\pi)_{\phi'_k}$, where the new phases $\phi'_j$ are called in the toggling frame and own the relation with original phases $\phi_j$ as

$$\phi'_j = (-1)^{j+1}\phi_j + \sum_{k<j}(-1)^{k+1}2\phi_k. \tag{6}$$

From this relation and Eq. (3) we can obtain a basic constraint on $\phi'_j$ as

$$2\sum_{j=1}^{n/2}(\phi'_{2j} - \phi'_{2j-1}) = -\Phi. \tag{7}$$

Then the conditions that the toggling frame phases $(\phi')$ must satisfy to remove the first- and second-order PSE have been derived as

$$\sum\cos(\phi'_j) = \sum\sin(\phi'_j) = 0, \tag{8}$$

and
$$\sum_j \sum_{k<j} \sin(\phi'_j - \phi'_k) = 0, \tag{9}$$

respectively [27].

Similarly, the conditions for the error tolerance of ORE to second-order also have been provided

$$\sum \cos(\phi''_j) = \sum \sin(\phi''_j) = 0, \tag{10}$$

$$\sum_j \sum_{k<j} \sin(\phi''_j - \phi''_k) = 0, \tag{11}$$

where the modified phases $\phi''_j$ are given by

$$\phi''_j = (-1)^{j+1}(\pi/2) + \phi'_j. \tag{12}$$

Note Eq. (10) alone corresponds to removing the first-order ORE, while Eq. (11) together with Eq. (8) to the second-order.

With these basic conditions for correcting PSE and ORE in hand, we are now in a position to discuss how to design robust composite pulses in detail. In general, the different number of $\pi$ pulses or geometric configurations can lead to distinct error tolerances. To better comprehend the effects of various phase patterns $(\boldsymbol{\phi})$, we point out the concept of an *equivalence family* of solutions, which indicates a family of composite pulses that exhibit essentially the same performances in the presence of some sort of errors. For example, the result of offsetting all phase angles in a certain sequence $(\boldsymbol{\phi})$ by an arbitrary amount $\Delta$, i.e. $(\boldsymbol{\phi}+\Delta)$, exhibits the fundamentally same behavior. In the following discussions, we focus on seeking genuinely distinct classes of pulse sequences as desired solutions.

## 3. First-order corrections

The condition for correcting first-order PSE in Eq. (8) has a straightforward geometric interpretation that the unit vectors $\boldsymbol{n}'_j = (\cos\phi'_j, \sin\phi'_j)$ must form a closed equilateral polygon of even order $n$. We start from the simplest case of $n=4$, where the four pulses must form a rhombus in the toggling frame as shown in Fig. 2(a).

Considering the basic relations for the rhombus

$$\phi'_3 = \phi'_1 + \pi, \phi'_4 = \phi'_2 + \pi, \tag{13}$$

which combined with the condition in Eq. (7) gives

$$\phi'_2 - \phi'_1 = -\Phi/4. \tag{14}$$

With loss of generality assume $\phi'_1 = (-3\Phi + 4\pi)/8$ so we can find a solution as

$$(\boldsymbol{\phi}') = (\phi'_1, \phi'_2, \phi'_3, \phi'_4) = (-3\Phi + 4\pi, -5\Phi + 4\pi, -3\Phi + 12\pi, -5\Phi + 12\pi)/8, \tag{15}$$

and our choice here particularly makes $(\boldsymbol{\phi})$ an antisymmetric phase sequence

$$(\boldsymbol{\phi}) = (\phi_1, \phi_2, \phi_3, \phi_4) = (-3\Phi + 4\pi, -\Phi + 4\pi, \Phi + 12\pi, 3\Phi + 12\pi)/8. \tag{16}$$

Such a pulse sequence owns the fidelity

$$\mathcal{F}_\epsilon = 1 - \sin^2(\Phi/4)\pi^4\epsilon^4/8 + O(\epsilon^6). \tag{17}$$

Then, the fidelity for the solution in Eq. (16) in the presence of ORE can also be investigated as

$$\mathcal{F}_f = 1 - 2\sin^2(\Phi/4)f^4 + O(f^5), \tag{18}$$

which exactly corrects first-order ORE. Such a performance simply relies on the conversion between $(\boldsymbol{\phi}')$ and $(\boldsymbol{\phi}'')$ in Eq. (12), which clearly keeps the shape of $(\boldsymbol{\phi}'')$ also a rhombus that satisfies Eq. (10). The observation of such simultaneous error tolerance would be further explained in section 5.

Note that replacing one + sign in Eq. (13) with the minus sign corresponds to swapping the order of the unit vectors $(1', 3')$ or $(2', 4')$ in Fig. 2(a), which would result in the same fidelities. In contrast, another reordering of the indices, i.e. the case of an arrowhead which presents

$$\phi_2' = \phi_1' \pm \pi, \phi_4' = \phi_3' \pm \pi, \tag{19}$$

can't satisfy the basic relation as required in Eq. (7) at all.

Besides, the use of equivalent constraint $Z_\Phi = Z_{\Phi+2\pi}$ in Eq. (3) with the setting $\phi_1' = (-3\Phi - 10\pi)/8$ would give a new solution

$$(\boldsymbol{\phi}) = (-3\Phi - 10\pi, -\Phi - 6\pi, \Phi + 6\pi, 3\Phi + 10\pi)/8 \tag{20}$$

with fidelities as

$$\mathcal{F}_\epsilon = 1 - \cos^2(\Phi/4)\pi^4\epsilon^4/8 + O(\epsilon^6), \tag{21}$$

$$\mathcal{F}_f = 1 - 2\cos^2(\Phi/4)f^4 + O(f^5). \tag{22}$$

For brevity, in this paper we denote a proposed family of solution $(\boldsymbol{\phi})$ to implement robust $z$-axis rotations as $S^{(n)}_{(i,j,k)}$, where the superscript $n$ and subscript $(i, j, k)$ together indicate the $k$th kind of $n$ $\pi$-pulse sequence that can suppress the $i$th-order PSE and $j$th-order ORE. By this notation, the solutions presented in Eq. (16) and Eq. (20) are denoted $S^{(4)}_{(1,1,1)}$ and $S^{(4)}_{(1,1,2)}$ respectively, and their performances are shown in second and third rows of Fig. 1. As expected from comparing Eq. (17) ((18)) with (21) ((22)), for $\Phi = \pi$ the fidelity plots of $S^{(4)}_{(1,1,1)}$ and $S^{(4)}_{(1,1,2)}$ are symmetric to each other around the line $f=0$, while for $\Phi = \pi/2, \pi/4$ the former outperforms the latter regarding both PSE and ORE.

Although for QC we usually focus on the regions of small errors, it is also interesting to note the extensions of the four sharp corners of the inmost contour line in each fidelity plot. This observation indicates that such designed pulses can retain moderate robustness at certain large errors, and may bring us specific applications in other quantum control scenarios [40, 41].

## 4. Second-order corrections

Since the presented solutions for $n=4$ correct first-order PSE and ORE at best, we proceed to $n=6$ for handling higher-order errors, where the situation becomes more complicated due to more adjustable phase parameters in the pulse sequence. In this section we only study several typical cases $S_{(2,0,k)}$ which can correct second-order PSE but has no suppression of ORE, while other cases for the simultaneous error tolerances will be further investigated in section 5.

### 4.1. Two equilateral triangles

To correct the first-order PSE, six unit vectors describing the phases $(\boldsymbol{\phi}')$ obeying Eq. (8) could form two equilateral triangles denoted $(1', 2', 3')$ and $(4', 5', 6')$. Actually, their relations

$$\phi_1' - \phi_2' = \phi_2' - \phi_3' = \pm 2\pi/3; \tag{23}$$

$$\phi_4' - \phi_5' = \phi_5' - \phi_6' = \mp 2\pi/3 \tag{24}$$

also automatically satisfy Eq. (9), leading to the correction of second-order PSE. Such properties can only be observed for the conditions where the $\pm$ and $\mp$ signs are chosen with the opposite signs, i.e., $(1', 2', 3')$ rotating clockwise with $(4', 5', 6')$ rotating anticlockwise as shown in Fig. 2(b) or vice versa.

Combined with Eq. (7), we have a fixed relative angle between these two triangles $\phi_2' - \phi_5' = \Phi/2$. Without loss of generality, we can set the only free parameter as $\phi_5' = -\Phi$ and thus obtain a pair of equivalent solutions

$$\begin{aligned}(\boldsymbol{\phi}') &= (\phi_1', \phi_2', \phi_3', \phi_4', \phi_5', \phi_6') \\ &= (-3\Phi \pm 4\pi, -3\Phi, -3\Phi \mp 4\pi, -6\Phi \mp 4\pi, -6\Phi, -6\Phi \pm 4\pi)/6\end{aligned}, \tag{25}$$

$$\begin{aligned}(\boldsymbol{\phi}) &= (\phi_1, \phi_2, \phi_3, \phi_4, \phi_5, \phi_6) \\ &= (-3\Phi \pm 4\pi, -3\Phi \pm 8\pi, -3\Phi \pm 4\pi, \pm 4\pi, \pm 8\pi, \pm 4\pi)/6\end{aligned}, \tag{26}$$

which have the fidelity

$$\mathcal{F}_\epsilon = 1 - \cos^2(\Phi/4)\pi^6\epsilon^6/32 + O(\epsilon^8). \tag{27}$$

Once the geometry of the phase pattern in the toggling frame $(\boldsymbol{\phi}')$ is determined, then the error tolerance of the sequence would subtly depend on the index arrangements for each unit vector $\boldsymbol{n}_j'$. For example, the feature of the discussed grouping indices $(1', 2', 3')$ $(4', 5', 6')$ for two equilateral triangles lies in characterizing one group by three adjacent toggling frame phases and the other group by three others. This pattern also applies to the set of indices $(2', 3', 4')$ $(1', 5', 6')$ and $(3', 4', 5')$ $(1', 2', 6')$, which are indeed equivalence solutions under cyclic reorderings of the component phases in the pulse sequence (cf. Ref. [24]).

In total there are $C_6^3/2 = 10$ grouping sets of indices for marking two equilateral triangles, which can be classified into three categories according to their performances:

(a) $(1', 2', 3')$ $(4', 5', 6')$, $(2', 3', 4')$ $(1', 5', 6')$, $(3', 4', 5')$ $(1', 2', 6')$;

(b) $(1', 2', 5')$ $(3', 4', 6')$, $(2', 3', 6')$ $(1', 4', 5')$, $(1', 3', 4')$ $(2', 5', 6')$, $(1', 3', 6')$ $(2', 4', 5')$, $(1', 2', 4')$ $(3', 5', 6')$, $(1', 4', 6')$ $(2', 3', 5')$;

(c) $(1', 3', 5')$ $(2', 4', 6')$.

Here, the class (a) has been discussed and can be denoted by $S^{(6)}_{(2,0,1)}$; (b) can only correct first-order PSE by direct verifications; and (c) owns simultaneous error tolerance of ORE and PSE, which will be further discussed in section 5.

### 4.2. Three pairs of antiparallel vectors

When considering three pairs of antiparallel vectors of six toggling frame phases that can remove first-order PSE, there are totally $C_6^2 C_4^2 / 3! = 15$ grouping sets of indices categorized into three classes:

(a). $(1',2')\,(3',4')\,(5',6')$, $(2',3')\,(4',5')\,(1',6')$, $(1',4')\,(2',5')\,(3',6')$, $(1',2')\,(3',6')\,(4',5')$, $(2',3')\,(5',6')\,(1',4')$, $(2',5')\,(3',4')\,(1',6')$;

(b). $(1',2')\,(3',5')\,(4',6')$, $(2',3')\,(4',6')\,(1',5')$, $(3',4')\,(2',6')\,(1',5')$, $(4',5')\,(2',6')\,(1',3')$, $(1',3')\,(2',4')\,(5',6')$, $(2',4')\,(1',6')\,(3',5')$;

(c). $(1',3')\,(2',5')\,(4',6')$, $(2',4')\,(1',5')\,(3',6')$, $(3',5')\,(2',6')\,(1',4')$.

It can be easily verified that class (a) doesn't satisfy the basic condition in Eq. (7), while class (b) can only correct the PSE to first-order. Therefore, we focus on the cases in class (c) to demonstrate their potential for correcting second-order PSE.

For example, the relations for $(1',3')\,(2',5')\,(4',6')$ as plotted in Fig. 2(c)
$$\phi'_3 - \phi'_1 = \pi,\ \phi'_5 - \phi'_2 = \pi,\ \phi'_6 - \phi'_4 = \pi, \tag{28}$$
which combined with Eq. (7) lead to the phases with two parameters $(\alpha, \beta)$
$$(\boldsymbol{\phi}') = (\alpha, \beta, \alpha+\pi, \alpha+\pi/2-\Phi/4, \beta+\pi, \alpha+3\pi/2-\Phi/4). \tag{29}$$
Solving for the cases that can realize Eq. (9) gives $\alpha - \beta = (k+1/4)\pi + \Phi/8$ with an integer $k$. By choosing $k=0$ and $\beta = -\Phi/8$ we have the solution
$$(\boldsymbol{\phi}) = (\pi/4, \pi/2+\Phi/8, -\pi/4+\Phi/4, \pi/4+\Phi/2, \pi/2+5\Phi/8, -\pi/4+3\Phi/4) \tag{30}$$
denoted as $S^{(6)}_{(2,0,2)}$. Also, other two grouping indices in class (c), i.e. $(2',4')\,(1',5')\,(3',6')$ and $(3',5')\,(2',6')\,(1',4')$ are equivalent to $S^{(6)}_{(2,0,2)}$. Besides, this solution family owns the same PSE fidelity as Eq. (27) by calculation.

### 4.3. Antisymmetric sequences

The use of symmetric and antisymmetric pulse sequences may open the possibility of new solutions [24, 42]. Since a symmetric sequence with even $n$ $\pi$ pulses can't satisfy the basic condition in Eq. (3), we therefore consider antisymmetric pulses to implement a $Z_\Phi$ gate robust to second-order PSE.

To analytically solve this situation, we might as well start from an antisymmetric toggling frame phases
$$(\boldsymbol{\phi}') = (\alpha, \beta, -\alpha+\beta+\Phi/4, \alpha-\beta-\Phi/4, -\beta, -\alpha) \tag{31}$$
as shown in Fig. 2(d) with corresponding

$$(\phi)=(\alpha, 2\alpha-\beta, \alpha-\beta+\Phi/4, -\alpha+\beta+3\Phi/4, -2\alpha+\beta+\Phi, -\alpha+\Phi), \tag{32}$$

which is obviously equivalent to an antisymmetric family

$$(\phi)=(\alpha-\Phi/2, 2\alpha-\beta-\Phi/2, \alpha-\beta-\Phi/4, -\alpha+\beta+\Phi/4, -2\alpha+\beta+\Phi/2, -\alpha+\Phi/2) \tag{33}$$

by offsetting all phase angles by $-\Phi/2$.

Since Eq. (31) automatically sets $\sum \sin(\phi'_j)$ to zero, we then seek parameters $(\alpha, \beta)$ that can satisfy both $\sum \cos(\phi'_j)=0$ and Eq. (9). It turns out such solutions exactly occur at $\beta=2\alpha-\Phi/4$ with choosing

$$\alpha = \frac{\Phi}{8} \pm \arccos\{\frac{1}{2}[-\cos(\Phi/8)+\sqrt{t_\Phi} \pm \sqrt{-t_\Phi+3\cos^2(\Phi/8)+\sin(\Phi/8)\sin(\Phi/4)/\sqrt{t_\Phi}}]\}, \tag{34}$$

with $t_\Phi = \cos^2(\Phi/8) + \cos^{1/3}(\Phi/4) + \cos^{2/3}(\Phi/4)$.

The two $\pm$ signs in Eq. (34) take the same value and indicate two distinct solutions denoted $S^{(6)}_{(2,0,3)}$ and $S^{(6)}_{(2,0,4)}$ with the same fidelities under only PSE as those of $S^{(6)}_{(2,0,1)}$ and $S^{(6)}_{(2,0,2)}$.

### 4.4. Off-resonance errors

Above proposed various solutions exhibit the same performance for PSE, but their effects on ORE are distinct. The solution $S^{(6)}_{(2,0,1)}$ has the fidelity

$$\mathcal{F}_f = 1 - 8\sin^2(\Phi/4)f^2 + O(f^4), \tag{35}$$

while the fidelity of $S^{(6)}_{(2,0,2)}$ is

$$\mathcal{F}_f = 1 - 2f^2 + O(f^4). \tag{36}$$

The comparison between Eqs. (35) and (36) indicates that for $\Phi=\pi (\pi/2, \pi/4)$, $S^{(6)}_{(2,0,2)}$ ($S^{(6)}_{(2,0,1)}$) has the better performance, which are confirmed in Fig. 3. Also, other two solutions $S^{(6)}_{(2,0,3)}$ and $S^{(6)}_{(2,0,4)}$ show similar thin regions enclosed by the inmost contour line as expected. These sharp fidelity regions of $S^{(6)}_{(2,0,k)}$ compared with $S^{(4)}_{(1,1,1)}$ and $S^{(4)}_{(1,1,2)}$ in Fig. 1 indicate that insensitivity to one type of error (PSE) may be obtained at the cost of increased sensitivity to another type of error (ORE). In the following, we explore the possibility of broadening the high-precision region in both dimensions.

### 5. Simultaneous error tolerance

In real physical experiments, the desirable simultaneous tolerance of PSE and ORE can be achieved by the thought of *separate cancellation* on toggling frame phases $(\phi')$ [27]. The solution $S^{(4)}_{(1,1,1)}$ for $n=4$ is such an example, and we proceed to the case $n=6$.

The constraints on $(\phi')$ are straightforward that the three odd phases $\phi'_1$, $\phi'_3$, and $\phi'_5$ differ by $\pm 2\pi/3$, and so do the even phases $\phi'_2$, $\phi'_4$, and $\phi'_6$. However, similar to the discussions in section 4.1, distinct families of solutions with respective robustness can actually arise by the choice of the $\pm$ signs in the constraints. For example, solving the case of the anticlockwise $(1',3',5')$ and clockwise $(2',4',6')$ or vice versa would give corresponding solutions as

$$(\phi') = (\alpha, \alpha - \Phi/6 \pm 4\pi/3, \alpha \pm 2\pi/3, \alpha - \Phi/6 \pm 2\pi/3, \alpha \pm 4\pi/3, \alpha - \Phi/6), \quad (37)$$

$$(\phi) = (\alpha, \alpha + \Phi/6 \pm 2\pi/3, \alpha + \Phi/3, \alpha + \Phi/2, \alpha + 2\Phi/3 \pm 2\pi/3, \alpha + 5\Phi/6). \quad (38)$$

Choosing $\alpha = 0$ leads to two equivalent pulse sequences in a compact formula

$$(\phi) = (0, \Phi/6 \pm 2\pi/3, \Phi/3, \Phi/2, 2\Phi/3 \pm 2\pi/3, 5\Phi/6) \quad (39)$$

denoted $S^{(6)}_{(2,2,1)}$, which own the fidelities in the presence of PSE and ORE as

$$\mathcal{F}_\epsilon = 1 - \cos^2(\Phi/4)\pi^6 \epsilon^6/32 + O(\epsilon^8), \quad (40)$$

$$\mathcal{F}_f = 1 - 2\sin^2(\Phi/4) f^6 + O(f^7). \quad (41)$$

Notice that offsetting phases of $(2',4',6')$ in Eq. (37) by $\mp \pi/3$ would lead to a new pair of solutions ($\alpha = 0$)

$$(\phi') = (0, -\Phi/6 \pm \pi, \pm 2\pi/3, -\Phi/6 \pm \pi/3, \pm 4\pi/3, -\Phi/6 \mp \pi/3), \quad (42)$$

$$(\phi) = (0, \Phi/6 \pm \pi, \Phi/3 \pm 2\pi/3, \Phi/2 \pm \pi, 2\Phi/3, 5\Phi/6 \mp \pi/3), \quad (43)$$

denoted $S^{(6)}_{(2,2,2)}$, which have the fidelities

$$\mathcal{F}_\epsilon = 1 - \sin^2(\Phi/4)\pi^6 \epsilon^6/32 + O(\epsilon^8), \quad (44)$$

and

$$\mathcal{F}_f = 1 - 2\cos^2(\Phi/4) f^6 + O(f^7), \quad (45)$$

respectively. In fact, these distinct solutions actually arise from the equivalent constraints $Z_\Phi = Z_{\Phi+2\pi} = Z_{\Phi-2\pi}$ used in Eq. (3), i.e. leading to $-\Phi \to -\Phi \mp 2\pi$ at the right hand side of Eq. (7). The behaviors of $S^{(6)}_{(2,2,1)}$ and $S^{(6)}_{(2,2,2)}$ are plotted in Fig. 4 in accordance with their analytic fidelity expressions. In each column, for $\Phi = \pi$ the two plots are symmetric around the line $f = 0$; for $\Phi = \pi/2, \pi/4$, $S^{(6)}_{(2,2,1)}$ has better performance for ORE while $S^{(6)}_{(2,2,2)}$ performs better for PSE. Besides, the cases when $(1',3',5')$ and $(2',4',6')$ rotate in the same direction can only simultaneously correct first-order PSE and ORE, and thus are not described here.

## 6. Sequences with eight pulses

We can straightforwardly extend above approach to longer sequences, e.g. the case $n=8$, for achieving higher-order simultaneous error tolerances. In this case, the four odd pulses $(1',3',5',7')$ and even pulses $(2',4',6',8')$ need to form either a rhombus or an arrowhead. For example, the case of two rhombuses, i.e. $\phi'_5 - \phi'_1 = \phi'_7 - \phi'_3 = \pi$ and $\phi'_6 - \phi'_2 = \phi'_8 - \phi'_4 = \pi$, together with the basic constraint in Eq. (7) lead to the solution with three free parameters as

$$(\phi') = (\alpha, \beta, \gamma, \alpha - \beta + \gamma - \Phi/4, \alpha + \pi, \beta + \pi, \gamma + \pi, \alpha - \beta + \gamma - \Phi/4 + \pi). \quad (46)$$

Then, Eqs. (9) and (11) need to be satisfied under appropriate $(\alpha, \beta, \gamma)$ for the simultaneous suppression of second-order PSE and ORE. An convenient technique is to solve a pair of equations $\sum_j \sum_{k<j} \sin(\phi'_j - \phi'_k) \pm \sin(\phi''_j - \phi''_k) = 0$, and we obtain $\beta = \alpha - \Phi/8 \pm \pi/2$, $\gamma = \alpha \pm \pi/2$. This pair of equivalent solutions with $\alpha = 0$ are

$$(\boldsymbol{\phi'}) = (0, -\frac{\Phi}{8} \pm \frac{\pi}{2}, \pm \frac{\pi}{2}, -\frac{\Phi}{8}, \pm \pi, -\frac{\Phi}{8} \pm \frac{3\pi}{2}, \pm \frac{3\pi}{2}, -\frac{\Phi}{8} \pm \pi), \tag{47}$$

$$(\boldsymbol{\phi}) = (0, \frac{\Phi}{8} \mp \frac{\pi}{2}, \frac{\Phi}{4} \mp \frac{\pi}{2}, \frac{3\Phi}{8}, \pm \pi + \frac{\Phi}{2}, \frac{5\Phi}{8} \pm \frac{\pi}{2}, \frac{3\Phi}{4} \pm \frac{\pi}{2}, \frac{7\Phi}{8} \pm \pi), \tag{48}$$

which we denote $S^{(8)}_{(3,3,1)}$ with fidelities

$$\mathcal{F}_\epsilon = 1 - \sin^2(\Phi/4) \pi^8 \epsilon^8 / 128 + O(\epsilon^{10}), \tag{49}$$

$$\mathcal{F}_f = 1 - 2\sin^2(\Phi/4) f^8 + O(f^9). \tag{50}$$

Eq. (47) indicates the two squares $(1', 3', 5', 7')$ and $(2', 4', 6', 8')$ must be traced in opposite directions similar to that for $n=6$ in Eq. (37),

Note that offsetting phases of $(2', 4', 6', 8')$ in Eq. (47) by $\mp \pi/4$ would lead to a pair of new solutions

$$(\boldsymbol{\phi}) = (0, \Phi/8 \mp \pi/4, \Phi/4, 3\Phi/8 \pm 3\pi/4, \Phi/2, 5\Phi/8 \pm 7\pi/4, 3\Phi/4, 7\Phi/8 \pm 11\pi/4) \tag{51}$$

denoted $S^{(8)}_{(3,3,2)}$, which own the fidelities

$$\mathcal{F}_\epsilon = 1 - \cos^2(\Phi/4) \pi^8 \epsilon^8 / 128 + O(\epsilon^{10}), \tag{52}$$

$$\mathcal{F}_f = 1 - 2\cos^2(\Phi/4) f^8 + O(f^9). \tag{53}$$

The performances of $S^{(8)}_{(3,3,1)}$ and $S^{(8)}_{(3,3,2)}$ are plotted in Fig. 5, and the comparison between them is quite similar to that between $S^{(4)}_{(1,1,1)}$ and $S^{(4)}_{(1,1,2)}$ shown in Fig. 1. Besides, the cases where the four odd or even togging frame pulses form an arrowhead have poor performances and are not further presented.

## 7. Comparison with other approaches

Now we compare our solutions with those previously proposed composite pulses for robust $Z_\Phi$ gates, and here the operation time cost of a sequence is defined as $T = \sum_i |\theta_i|/\pi$ with the $i$th rotation angle $|\theta_i|$ [38]. For example, for $\Phi = \pi$ the robust $Z_\Phi$ gate constructed by Ichikawa *et al.* [38] has an explicit form as $(\pi)_0 (2\pi)_{3.566} (2\pi)_{1.147} (\pi)_{-\pi/2}$ with $T=6$, and its fidelity $\mathcal{F}_\epsilon = 1 - 8.24\epsilon^4 + O(\epsilon^6)$ indicates first-order PSE correction. Besides, for achieving simultaneous error tolerance, the nested composite pulse of their solution is $(\pi/3)_0 (5\pi/3)_\pi (7\pi/3)_0 (2\pi)_{3.566} (2\pi)_{1.147} (\pi/3)_{-\pi/2} (5\pi/3)_{-3\pi/2} (7\pi/3)_{-\pi/2}$ with $T=12.7$, which can simultaneously suppress first-order PSE and ORE. As comparison, our constructed solution $S^{(4)}_{(1,1,1)}$ with just $T=4$ can achieve such a performance of first-order correction.

Other existing composite pulses only robust against PSE, i.e. the SCROFULOOUS [20], SK1 [28] and BB1 [43] pulses, are also considered. By direct verification, the SCROFULOUS sequence $Z_\Phi = (\pi)_{\pi/3}(\pi)_{5\pi/3}(\pi)_{\pi/3}(\pi)_{\pi/3-\Phi/2}(\pi)_{5/3-\Phi/2}(\pi)_{\pi/3-\Phi/2}$ is indeed equivalent to our proposed solution $S^{(6)}_{(2,0,1)}$ with $T=6$ in Eq. (26), while the fidelities $\mathcal{F}_\epsilon$ of SK1 sequence $Z_\Phi = (2\pi)_\varphi(2\pi)_{-\varphi}(\pi)_0(2\pi)_{\varphi-\Phi/2}(2\pi)_{-\varphi-\Phi/2}(\pi)_{-\Phi/2}$ with $\varphi = \arccos(-1/4)$ and $T=10$ are slightly worse than $S^{(6)}_{(2,0,1)}$, and the performance of BB1 sequence $Z_\Phi = (\pi)_\varphi(2\pi)_{3\varphi}(\pi)_\varphi(\pi)_0(\pi)_{\varphi-\Phi/2}(2\pi)_{3\varphi-\Phi/2}(\pi)_{\varphi-\Phi/2}(\pi)_{-\Phi/2}$ with $T=10$ under PSE is comparable to our solution $S^{(6)}_{(2,2,2)}$ with $T=6$.

For simultaneous error tolerance, various concatenated composite pulses (CCCPs) have been designed, including the reduced CinSK (CORPSE in SK1), reduced CinBB (CORPSE in BB1) and reduced SKinsC (SK1 in short CORPSE) sequences [29] with $T=16.7$, $T=16.7$ and $T=12.7$ for implementing robust $Z_\Phi$ gates, respectively [38]. Since these CCCPs can only correct PSE to second order and ORE to first order, our solutions $S^{(8)}_{(3,3,1)}$ and $S^{(8)}_{(3,3,2)}$ with $T=8$ outperform them in terms of time cost and error tolerance. In summary, the $Z_\Phi$ gate can be implemented with a shorter time and better robustness by our construction, in comparison with certain existing approaches.

## 8. Arbitrarily accurate $\pi$-pulse sequences

Inspired by the toggling frame phases $(\phi')$ of solutions for the cases $n=4$, 6 and 8, arbitrarily accurate $\pi$ composite pulses are derived here by simple formulas, which can correct PSE and ORE to any desired order simultaneously.

For the case $n=4$, the toggling phases $(1', 3')$ and $(2', 4')$ must form two pairs of antiparallel vectors for deriving two distinct solutions $S^{(4)}_{(1,1,1)}$ and $S^{(4)}_{(1,1,2)}$ in section 3. For $n=6$, the toggling phases in Eqs. (37) and (42) for solutions $S^{(6)}_{(2,2,1)}$ and $S^{(6)}_{(2,2,2)}$ can be generalized to such families of solutions

$$(\phi'_1, \phi'_3, \phi'_5) = (\alpha, \alpha \pm 2\pi/3, \alpha \pm 4\pi/3) + j_1\pi/3, \tag{54}$$

$$(\phi'_2, \phi'_4, \phi'_6) = (\alpha - \Phi/6, \alpha - \Phi/6 \mp 2\pi/3, \alpha - \Phi/6 \mp 4\pi/3) + j_2\pi/3. \tag{55}$$

The fidelities of these phase patterns are the same as Eqs. (40) and (41) for even $(j_1 - j_2)$, and as Eqs. (44) and (45) for odd $(j_1 - j_2)$.

Similarly, the patterns for $S^{(8)}_{(3,3,1)}$ and $S^{(8)}_{(3,3,2)}$ with $n=8$ can be generalized to

$$(\phi'_1, \phi'_3, \phi'_5, \phi'_7) = (\alpha, \alpha \pm \pi/2, \alpha \pm \pi, \alpha \pm 3\pi/2) + j_1\pi/4, \tag{56}$$

$$(\phi'_2, \phi'_4, \phi'_6, \phi'_8) = (\alpha - \Phi/8, \alpha - \Phi/8 \mp \pi/2, \alpha - \Phi/8 \mp \pi, \alpha - \Phi/8 \mp 3\pi/2) + j_2\pi/4. \tag{57}$$

Note the solutions for even (odd) integer $(j_1 - j_2)$ are equivalent to $S^{(8)}_{(3,3,1)}$ ($S^{(8)}_{(3,3,2)}$). Here, the additional integers $(j_1, j_2)$ actually corresponds to an equivalent modification of the right hand side of Eq. (7) by multiples of $2\pi$.

By observing these patterns for $n=4$, 6, 8 respectively, we heuristically present a sequence with even $n$ $\pi$ pulses which can simultaneously suppress the PSE and ORE to the order $O(\epsilon^n)$ and $O(f^n)$ in fidelities $\mathcal{F}_\epsilon$ and $\mathcal{F}_f$: the odd phases $(1', 3', \ldots, (n-1)')$ and even $(2', 4', \ldots, n')$ must form two $n/2$-edged closed regular polygons that are traced in opposite directions, and cases for $n \geq 6$ are classified into two categories for discussion.

(1). For $n = 4k+2$ ($k=1,2,3...$), the odd pulses in the toggling frame are chosen as

$$(\phi'_1, \phi'_3, ..., \phi'_{n-1}) = (\alpha, \alpha \pm 2\pi/(2k+1), ..., \alpha \pm 4k\pi/(2k+1)) + j_1\pi/(2k+1) \tag{58}$$

and associated even pulses are

$$(\phi'_2, \phi'_4, ..., \phi'_n) = (\alpha - \Phi/n, \alpha - \Phi/n \mp 2\pi/(2k+1), ..., \alpha - \Phi/n \mp (4k\pi/(2k+1))) + j_2\pi/(2k+1), \tag{59}$$

where the values of $\alpha$ and integers $j_1$, $j_2$ are arbitrary. The solution for the sign $+(-)$ in Eq. (58) together with $-(+)$ in Eq. (59) corresponds to one regular polygon along clockwise (anti-clockwise) with another along anti-clockwise (clockwise).

By direct verifications, the fidelities for the setting $(\alpha, j_1, j_2)$ can be summarized into

$$\mathcal{F}_\epsilon^{(n=4k+2)} = 1 - \frac{[1+(-1)^{j_1-j_2}\cos(\Phi/2)]}{2^n}(\pi\epsilon)^n + O(\epsilon^{n+2}), \tag{60}$$

$$\mathcal{F}_f^{(n=4k+2)} = 1 - [1+(-1)^{j_1-j_2+1}\cos(\Phi/2)]f^n + O(f^{n+1}). \tag{61}$$

(2). For the case $n=4k$ ($k=2,3...$), the toggling frame pulses are chosen as

$$(\phi'_1, \phi'_3, \phi'_5, ..., \phi'_{n-1}) = (\alpha, \alpha \pm \pi/k, \alpha \pm 2\pi/k, ..., \alpha \pm (2k-1)\pi/k) + j_1\pi/2k, \tag{62}$$

$$(\phi'_2, \phi'_4, \phi'_6, ..., \phi'_n)$$
$$= (\alpha - \Phi/n, \alpha - \Phi/n \mp \pi/k, \alpha - \Phi/n \mp 2\pi/k, ..., \alpha - \Phi/n \mp (2k-1)\pi/k) + j_2\pi/2k \tag{63}$$

The fidelities of these patterns can be summarized into

$$\mathcal{F}_\epsilon^{(n=4k)} = 1 - \frac{[1+(-1)^{j_1-j_2+1}\cos(\Phi/2)]}{2^n}(\pi\epsilon)^n + O(\epsilon^{n+2}), \tag{64}$$

$$\mathcal{F}_f^{(n=4k)} = 1 - [1+(-1)^{j_1-j_2+1}\cos(\Phi/2)]f^n + O(f^{n+1}). \tag{65}$$

For the rotational angle $\Phi \in (0, \pi]$, clearly the sequence with even ($j_1 - j_2$) own better fidelities.

For simplicity, in these cases we only list the patterns of toggling frame ($\phi'$), as the corresponding solution ($\phi$) can be obtained by Eq. (6). The above expressions of fidelities indicate that the length (time cost) of our constructed robust sequences scale linearly with the correction order as $T = O(n)$. Thus for practical implementation of $Z_\Phi$ gates by our design, composite $\pi$-pulses can be chosen appropriately according to particular circumstances, e.g. the required error tolerance, the value of rotational angle $\Phi$ and the dominant error (PSE or ORE).

## 9. Conclusions

In this paper, we concentrate on how to design $\pi$-pulse sequences to suppress the systematic error terms, i.e. PSE and ORE, by. By comparison with other existing schemes, our solutions turn out to be remarkably effective in terms of operation time cost and achieved error tolerance. Furthermore, we have generalized such construction and heuristically derived arbitrarily accurate $\pi$-pulse sequences with a linear length scaling.

Previous prominent investigations [24, 27, 38, 39] enlighten us on the design of robust pulse sequences. Starting from geometric observations is a beneficial idea, and initial progress can be further supplemented by the analysis of symmetric/antisymmetric properties and algebraic methods to obtain new types of solutions (e.g. see section 4.3). In terms of achieving simultaneous error tolerance to any desired order, we propose simple and intuitive solutions for $Z_\Phi$ gates which are easy to design and implement. Also, considering other types of pulses [29] as well as introducing some analytic or numerical methods [39] might further extend the study of arbitrary *z*-axis rotations.

In future work, these robust *z*-rotation gates are potentially good candidates for improving the implementations of specific quantum algorithms [44, 45]. In a broader sense, our designed phase and $\pi/8$ gates combined with existing Hadamard [38] and controlled-NOT gates [23] form a specific discrete set of gates that can be used to approximate any unitary operation to good accuracy in principle [46], and thus offer potential for a variety of QC and QIP in NMR [26] or other physical systems [47-49]

## Acknowledgements

These two authors L.Z. and S.H.Z contributed equally to this work.

## ORCID iDs

Li Zhang: https://orcid.org/0000-0001-9917-4518

Shihao Zhang: https://orcid.org/0000-0001-7725-4779

## References

[1]   Aspuru-Guzik A and Walther P 2012 Photonic quantum simulators *Nat. Phys.* **8** 285-91

[2]   Wendin G 2017 *Rep. Prog. Phys.* **80** 106001

[3]   Lekitsch B, Weidt S, Fowler A G, Mølmer K, Devitt S J, Wunderlich C and Hensinger W K 2017 *Sci. Adv.* **3** e1601540

[4]   Xin T, Huang S, Lu S, Li K, Luo Z, Yin Z, Li J, Lu D, Long G and Zeng B 2018 *Sci. Bull.* **63** 17-23

[5]   Li Y and Benjamin S C 2017 *Phys. Rev. X* **7** 021050

[6]   Banchi L, Pancotti N and Bose S 2016 *npj Quantum Inf.* **2** 16019

[7]   Zahedinejad E, Ghosh J and Sanders B C 2016 *Phys. Rev. Appl.* **6** 054005

[8]   Gambetta J M, Chow J M and Steffen M 2017 *npj Quantum Inf.* **3** 2

[9]   Gottesman D 1998 *Phys. Rev. A* **57** 127

[10]     Takita M, Cross A W, Corcoles A D, Chow J M and Gambetta J M 2017 *Phys. Rev. Lett.* **119** 180501
[11]     Jones J A, Vedral V, Ekert A and Castagnoli G 2000 *Nature* **403** 869
[12]     Zhu S L, Wang Z D and Zanardi P 2005 *Phys. Rev. Lett.* **94** 100502
[13]     Zanardi P and Rasetti M 1999 *Phys. Lett. A* **264** 94-9
[14]     Feng G, Xu G and Long G 2013 *Phys. Rev. Lett.* **110** 190501
[15]     Xu Y *et al* 2018 *Phys. Rev. Lett.* **121** 110501
[16]     Kitaev A Y 2003 *Ann. Phys.* **303** 2-30
[17]     Yao X-C *et al* 2012 *Nature* **482** 489-94
[18]     Jones J A 2003 *Phil. Trans. R. Soc. A* **361** 1429-40
[19]     Jones J A 2003 *Phys. Rev. A* **67** 012317
[20]     Cummins H K, Llewellyn G and Jones J A 2003 *Phys. Rev. A* **67** 042308
[21]     Mc Hugh D and Twamley J 2005 *Phys. Rev. A* **71** 012327
[22]     Ichikawa T, Bando M, Kondo Y and Nakahara M 2011 *Phys. Rev. A* **84** 062311
[23]     Ichikawa T, Güngördü U, Bando M, Kondo Y and Nakahara M 2013 *Phys. Rev. A* **87** 022323
[24]     Husain S, Kawamura M and Jones J A 2013 *J. Magn. Reson.* **230** 145-54
[25]     Jones J A 2013 *Phys. Lett. A* **377** 2860-2
[26]     Jones J A 2011 *Prog. Nucl. Magn. Reson. Spectrosc.* **59** 91-120
[27]     Jones J A 2013 *Phys. Rev. A* **87** 052317
[28]     Brown K R, Harrow A W and Chuang I L 2004 *Phys. Rev. A* **70** 052318
[29]     Bando M, Ichikawa T, Kondo Y and Nakahara M 2013 *J. Phys. Soc. Jpn.* **82** 014004
[30]     Nielsen M A and Chuang I L 2011 *Quantum Computation and Quantum Information: 10th Anniversary Edition* (Cambridge University Press)
[31]     Bocharov A and Svore K M 2012 *Phys. Rev. Lett.* **109** 190501
[32]     Kliuchnikov V, Maslov D and Mosca M 2013 *Phys. Rev. Lett.* **110** 190502
[33]     Zhou X, Leung D W and Chuang I L 2000 *Phys. Rev. A* **62** 052316
[34]     Fowler A G, Mariantoni M, Martinis J M and Cleland A N 2012 *Phys. Rev. A* **86** 032324
[35]     Weinstein Y S 2014 *Phys. Rev. A* **89** 020301
[36]     Fowler A G, Stephens A M and Groszkowski P 2009 *Phys. Rev. A* **80** 052312
[37]     Bravyi S and Haah J 2012 *Phys. Rev. A* **86** 052329
[38]     Ichikawa T, Filgueiras J G, Bando M, Kondo Y, Nakahara M and Suter D 2014 *Phys. Rev. A* **90** 052330
[39]     Low G H, Yoder T J and Chuang I L 2014 *Phys. Rev. A* **89** 022341
[40]     Torosov B T and Vitanov N V 2018 *Phys. Rev. A* **97** 043408
[41]     Torosov B T and Vitanov N V 2019 *Phys. Rev. A* **99** 013402
[42]     Odedra S, Thrippleton M J and Wimperis S 2012 *J. Magn. Reson.* **225** 81-92
[43]     Wimperis S 1994 *J. Magn. Reson., Ser. A* **109** 221-31

[44] Cummins H and Jones J 2000 *New J. Phys.* **2** 6
[45] Xiao L and Jones J A *Phys. Rev. A* **73** 032334
[46] Dawson C M and Nielsen M A 2006 *Quantum Inf. Comput.* **6** 81-95
[47] Timoney N, Elman V, Glaser S, Weiss C, Johanning M, Neuhauser W and Wunderlich C 2008 *Phys. Rev. A* **77** 052334
[48] Wang X, Bishop L S, Kestner J P, Barnes E, Sun K and Das Sarma S 2012 *Nat. Commun.* **3** 997
[49] Genov G T, Schraft D, Halfmann T and Vitanov N V 2014 *Phys. Rev. Lett.* **113** 043001

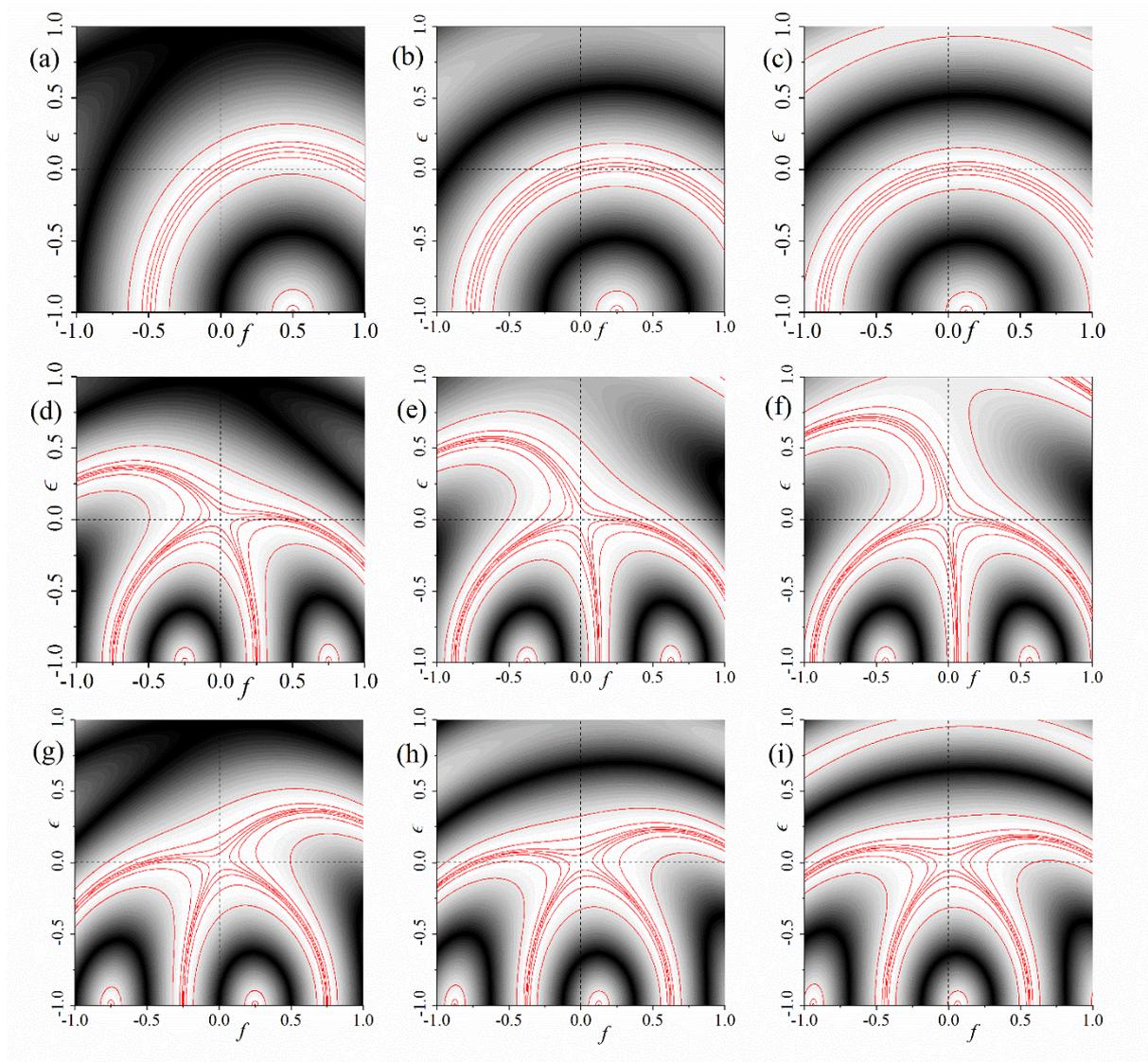

**Figure 1.** (Color online) Fidelities achieved by (a)(b)(c) baseline solutions with $n=2$, (d)(e)(f) the composite pulse solutions $S^{(4)}_{(1,1,1)}$ and (g)(h)(i) $S^{(4)}_{(1,1,2)}$ with $n=4$ giving simultaneous suppression of first-order PSE and ORE. In each row, the three plots indicate the case $\Phi = \pi, \pi/2$ and $\pi/4$, respectively. For $n=2$, contours are drawn at 0.9, 0.99, and 0.999 fidelity, while such logarithmically spaced infidelities continue for $n=4$ with the inmost contour at an infidelity of $10^{-4}$. The intersection of the horizontal and vertical dashed lines marks the center of the plot $(\epsilon, f) = (0,0)$.

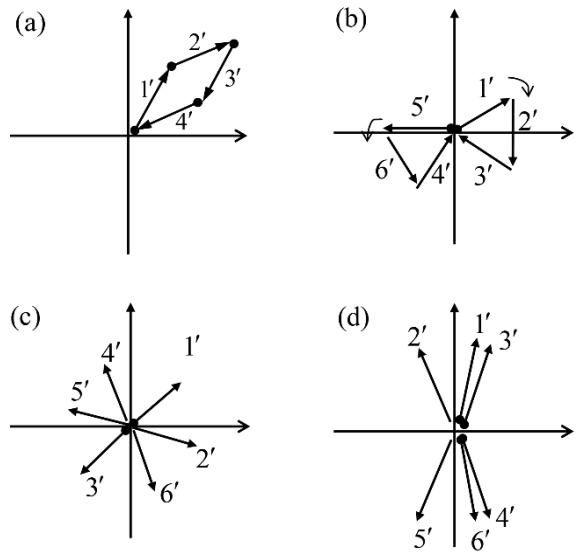

**Figure 2.** Vector diagrams describing the toggling frame phases for designed composite pulse sequences: (a) the case $n=4$ ; (b) (c) and (d) the cases presented in sections 4.1, 4.2 and 4.3 with $n=6$, respectively.

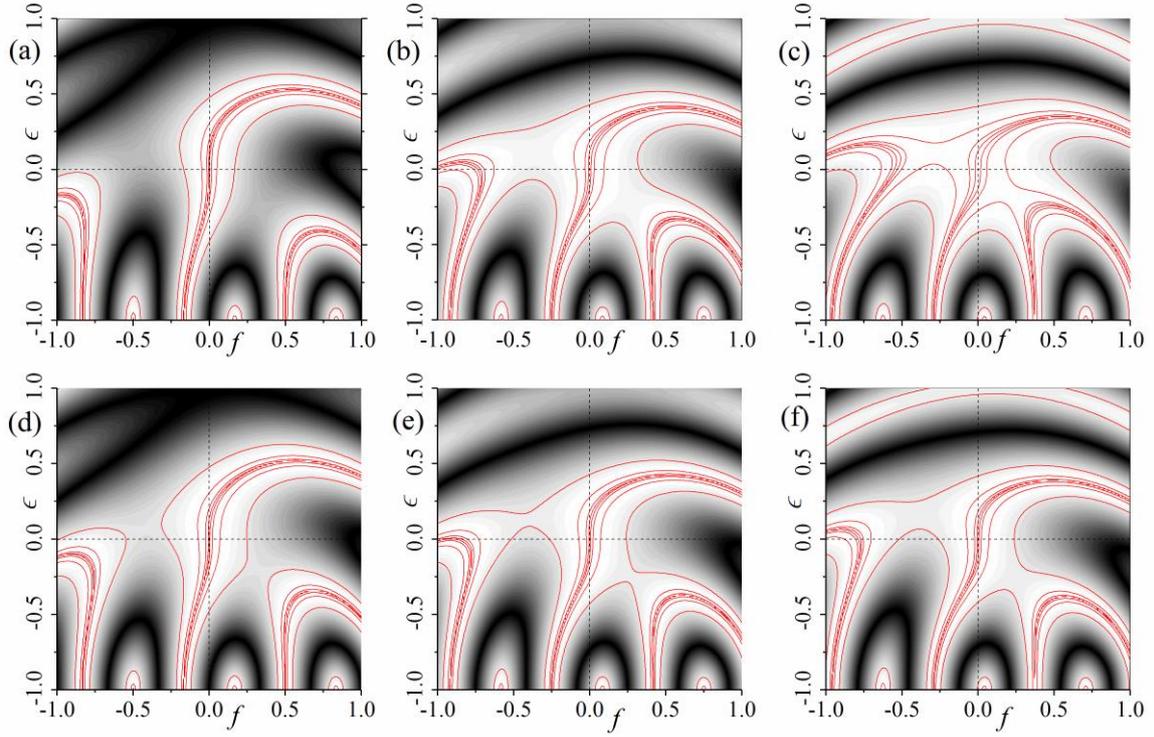

**Figure 3.** (Color online) Fidelities achieved by some composite pulses with *n*=6 designed to only suppress PSE to second order: (a)(b)(c) the $S^{(6)}_{(2,0,1)}$ pulse, (d)(e)(f) $S^{(6)}_{(2,0,2)}$ pulse. In each row, the three plots indicate the case $\Phi = \pi, \pi/2$ and $\pi/4$, respectively. The inmost contour is drawn at an infidelity of $10^{-4}$.

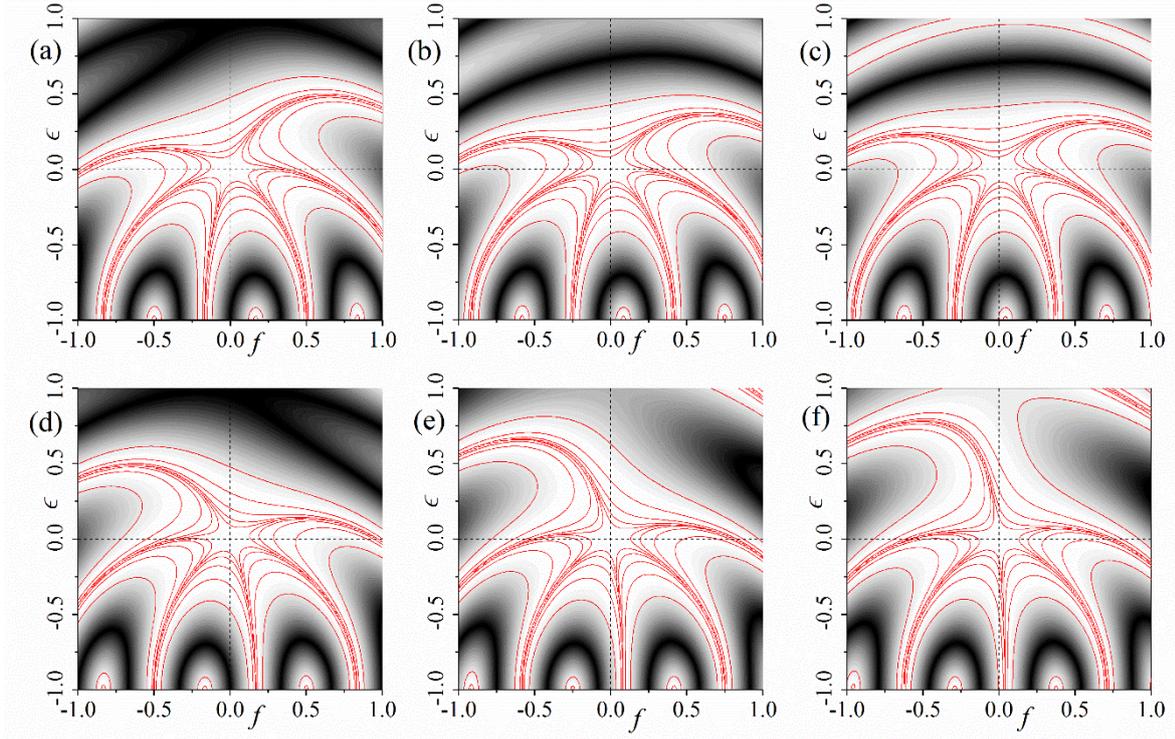

**Figure 4.** (Color online) Fidelities achieved by some composite pulses with *n*=6 designed to suppress both PSE and ORE to second order: (a)(b)(c) the $S^{(6)}_{(2,2,1)}$ pulse, (d)(e)(f) $S^{(6)}_{(2,2,2)}$ pulse. In each row, the three plots indicate the case $\Phi = \pi, \pi/2$ and $\pi/4$, respectively. The inmost contour is drawn at an infidelity of $10^{-5}$.

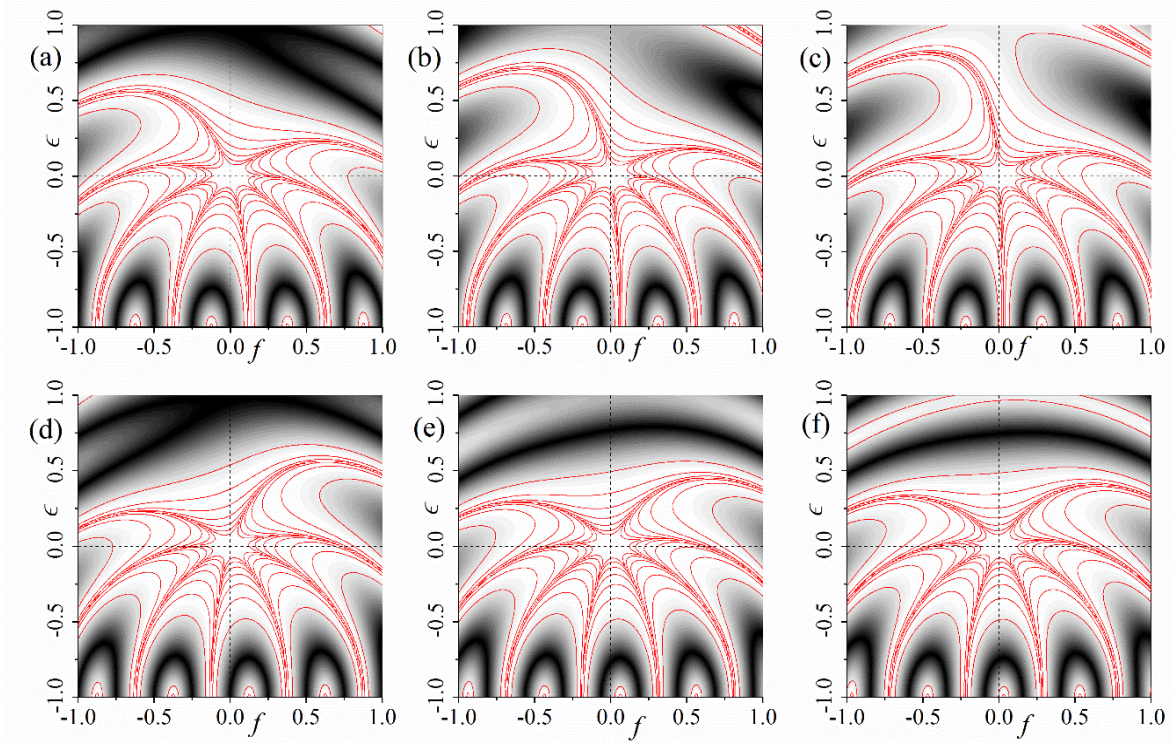

**Figure 5.** (Color online) Fidelities achieved by some composite pulses with *n*=8 designed to suppress both PSE and ORE to third order: (a)(b)(c) the $S^{(8)}_{(3,3,1)}$ pulse, (d)(e)(f) $S^{(8)}_{(3,3,2)}$ pulse. In each row, the three plots indicate the case $\Phi = \pi, \pi/2$ and $\pi/4$, respectively. The inmost contour is drawn at an infidelity of $10^{-7}$.